\documentclass[]{aa}
\usepackage{epsfig}
\def\gsim{\vcenter{\hbox{$>$}\offinterlineskip\hbox{$\sim$}}}
\begin{document}
\thesaurus{06(08.03.1; 08.03.4; 08.13.2; 08.16.4; 11.13.1; 13.09.6)}
\title{Luminous carbon stars in the Magellanic Clouds\thanks{This paper is
       based on data obtained at the Cerro Tololo Inter-american
       Observatories, Chile.}}
\author{Jacco Th. van Loon\inst{1}, Albert A. Zijlstra\inst{2} \and M.A.T.
        Groenewegen\inst{3}}
\institute{Astronomical Institute, University of Amsterdam, Kruislaan 403,
           NL-1098 SJ Amsterdam, The Netherlands
      \and University of Manchester Institute of Science and Technology,
           P.O.Box 88, Manchester M60 1QD, United Kingdom
      \and Max-Planck Institut f\"{u}r Astrophysik, Karl-Schwarzschild
           Stra{\ss}e 1, D-85740 Garching bei M\"{u}nchen, Germany}
\date{Received date; accepted date}
\maketitle
\markboth{Jacco Th.\ van Loon et al.:
          Luminous carbon stars in the Magellanic Clouds}
         {Jacco Th.\ van Loon et al.:
          Luminous carbon stars in the Magellanic Clouds}
\begin{abstract}

We present ground-based 3 $\mu$m spectra of obscured Asymptotic Giant Branch
(AGB) stars in the Magellanic Clouds (MCs). We identify the carbon stars on
the basis of the 3.1 $\mu$m absorption by HCN and C$_2$H$_2$ molecules.

We show evidence for the existence of carbon stars up to the highest AGB
luminosities ($M_{\rm bol}=-7$ mag, for a distance modulus to the LMC of 18.7
mag). This proves that Hot Bottom Burning (HBB) cannot, in itself, prevent
massive AGB stars from becoming carbon star before leaving the AGB. It also
sets an upper limit to the distance modulus of the Large Magellanic Cloud of
18.8 mag.

The equivalent width of the absorption band decreases with redder $(K-L)$
colour when the dust continuum emission becomes stronger than the photospheric
emission. Carbon stars with similar $(K-L)$ appear to have equally strong 3
$\mu$m absorption in the MCs and the Milky Way. We discuss the implications
for the carbon and nitrogen enrichment of the stellar photosphere of carbon
stars.

\end{abstract}

\section{Introduction}

As intermediate-mass stars ascend the Asymptotic Giant Branch (AGB), the
combination of a convective mantle with the occurrence of thermal pulses in
the interior of the star causes nuclear burning products to be transported to
the photosphere of the star ($3^{rd}$ dredge-up). It may enhance the carbon
over oxygen ratio in the photosphere. When this ratio exceeds unity, a carbon
star is born.

Two decades ago it was realised that in the Large Magellanic Cloud (LMC),
where reliable luminosities for individual stars could be obtained, no carbon
stars were known more luminous than $M_{\rm bol}\sim-6$ mag, whereas the AGB
extends up to $M_{\rm bol}\sim-7$ mag (e.g.\ Iben 1981). This was surprising,
as the lower metallicity in the LMC was thought to increase the efficiency of
$3^{rd}$ dredge-up. Recent studies confirm the lack of optically bright carbon
stars in the LMC (Costa \& Frogel 1996). Three solutions were suggested: (1)
high mass-loss rates yield a cut-off to the AGB evolution at $M_{\rm
bol}\sim-6$ mag, (2) nuclear burning at the bottom of the convective mantle of
massive AGB stars (Hot Bottom Burning, or HBB) prevents the dredge-up of
carbon (Iben \& Renzini 1983; Wood et al.\ 1983), or (3) luminous carbon stars
have become obscured by their dusty circumstellar envelope (CSE). The first
solution is ruled out by the detection in the LMC of Li-rich stars with
luminosities between $M_{\rm bol}=-6$ and $-7.2$ mag (Smith et al.\ 1995) and
AGB star luminosities in clusters in the LMC (Aaronson \& Mould 1985;
Westerlund et al.\ 1991). HBB has been commonly accepted as the solution
(Boothroyd et al.\ 1993).

We have investigated the possibility that massive carbon stars are
dust-enshrouded, by searching for obscured AGB counterparts of IRAS point
sources in the LMC and SMC, (Loup et al.\ 1997; Zijlstra et al.\ 1996; van
Loon et al.\ 1997, 1998a: papers I to IV). Chemical types could only be
identified for a few stars. Hence the luminosity distributions of obscured
oxygen-rich and carbon stars in the LMC are not well defined. One very
luminous carbon star candidate was found in the LMC, the obscured AGB star
IRAS04496$-$6958 (paper IV). Remarkably, Trams et al.\ (1999a,b) recently
presented ISO spectra showing that the CSE of this object also contains an
oxygen-rich dust component.

In this paper, we present ground-based L-band (3 $\mu$m) spectra of obscured
AGB stars in the Magellanic Clouds. We had the opportunity to do these
observations as a result of an exchange of observing time between the European
Southern Observatory (ESO) and the Cerro Tololo Inter-american Observatories
(CTIO). Absorption between 3.0 and 3.3 $\mu$m is due to HCN and C$_2$H$_2$
molecules in the extended atmospheres of carbon stars, whilst oxygen-rich,
M-type, stars display a featureless continuum at this wavelength (Merrill \&
Stein, 1976a,b,c; Noguchi et al.\ 1977; Ridgway et al.\ 1978). We show the
spectrum of IRAS04496$-$6958, confirming its carbon star nature. We also use
the 3 $\mu$m spectra of other luminous obscured AGB stars in the LMC and SMC
to classify them as carbon or M star. These are the first 3 $\mu$m spectra of
extra-galactic stars to be presented in the literature.

\section{Observations}

%
%
\begin{table}
\caption[]{Observing log of 3 $\mu$m spectroscopy of obscured AGB stars in the
Magellanic Clouds. We list IRAS and LI (Schwering \& Israel 1990)
designations, and alternative names. IRAS00393$-$7326 is a Faint Source
Catalogue entry, IRAS05203$-$6638 is from Reid (1991). Julian Dates are
$RJD=JD-2,450,000$.}
\begin{tabular}{llll}
\hline\hline
IRAS         & LI      & other   & RJD \\
\hline
00393$-$7326 &         &         & 821 \\
00554$-$7351 & SMC0119 &         & 821 \\
04496$-$6958 & LMC0057 &         & 441 \\
04498$-$6842 & LMC0060 &         & 820 \\
04509$-$6922 & LMC0077 &         & 441 \\
04553$-$6825 & LMC0181 & WOH G64 & 820 \\
05003$-$6712 & LMC0297 &         & 821 \\
05009$-$6616 & LMC0310 &         & 821 \\
05112$-$6755 & LMC0570 & TRM  4  & 441 \\
05128$-$6455 & LMC1880 &         & 821 \\
05203$-$6638 &         & TRM 88  & 820 \\
05291$-$6700 & LMC1137 &         & 820 \\
05298$-$6957 & LMC1164 &         & 441 \\
05300$-$6651 & LMC1177 & TRM 79  & 441 \\
05329$-$6708 & LMC1286 & TRM 60  & 821 \\
06028$-$6722 & LMC1817 &         & 820 \\
\hline
\end{tabular}
\end{table}

We used the Infra-Red Spectrometer (IRS) at the 4 m Blanco telescope at CTIO,
Chile, in December 1996 and January 1998 to take L-band spectra of a sample of
obscured AGB stars in the LMC (Table 1). We also included a red supergiant
(RSG) in the LMC (IRAS04553$-$6825) and two obscured AGB stars in the SMC
(IRAS00393$-$7326 and IRAS00554$-$7351).

The IRS uses a $256\times256$ InSb array, and is operated at the
f/30 focus. The slit length is $16^{\prime\prime}$ at a spatial scale of
$0.32^{\prime\prime}$\,pix$^{-1}$. We used the 75 lines mm$^{-1}$ grating in
$1^{st}$ order, covering 0.33 $\mu$m (the grating is blazed at 4.65 $\mu$m).
We observed with the grating centred at 3.04 and 3.34 $\mu$m to cover the
spectral region between 2.87 and 3.50 $\mu$m (IRAS00393$-$7326 was observed
with the grating centred at 3.04 $\mu$m only). The slit aperture was 0.35 mm
($0.7^{\prime\prime}$ on the sky), yielding a resolving power of $R\sim1300$.

We performed sequences with the star alternatingly placed at either of two
positions on the slit separated by $6^{\prime\prime}$ to facilitate background
subtraction. The frames that were saved on disc consisted of 10 coadded frames
of 2 s integration time each at 3.04 $\mu$m, and 20 coadded frames of 1 s at
3.34 $\mu$m. In December 1996 we used integration times of 3 and 1.5 s,
respectively. Total on-source integration times were typically eight minutes,
but observing times amount to at least five times as much. The stars of our
sample represent the faintest stars of which the continuum can be detected by
the IRS ($8^{th}$ mag in L-band), and that can be acquired through a
$K^{\prime}$ filter (limiting magnitude $K^{\prime}\sim12$ mag). The standard
stars HR77 ($\zeta$ Tuc) for the SMC and HR2015 ($\delta$ Dor) for the LMC
were observed regularly, with the star placed at five positions on the slit,
separated by $3^{\prime\prime}$. Weather conditions were non-photometric, with
thin cirrus passing over on the December and January 6/7 nights. Seeing was
typically $1^{\prime\prime}$, 0.7 to $1^{\prime\prime}$, and 1 to
$1.5^{\prime\prime}$. Atmospheric stability was least on January 7/8.

%
%
\begin{figure}[tb]
\centerline{\psfig{figure=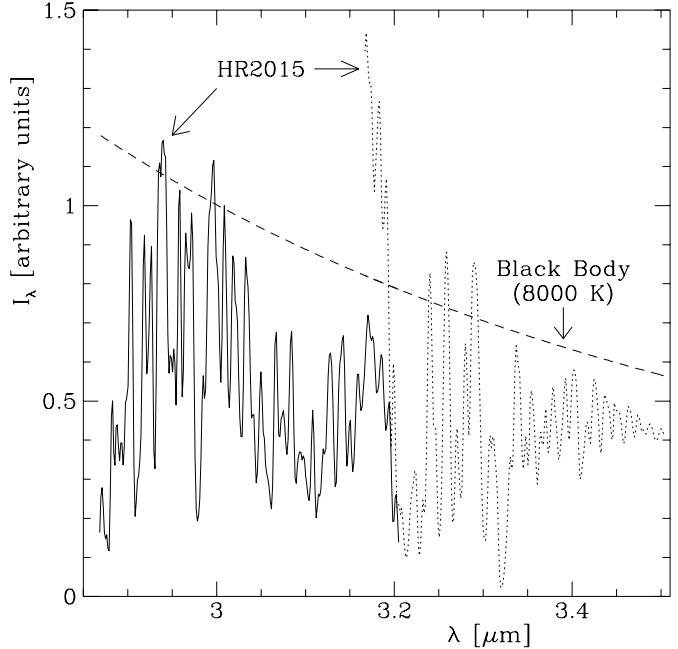,width=88mm}}
\caption[]{Composite spectrum of the standard star HR2015, with an 8000 K
blackbody for comparison. Telluric absorption is very important, leaving only
half of the flux to reach the detector. The instrument response function is
higher at the short-wavelength end than at the long-wavelength end, which is
why the two spectral settings do not match perfectly around 3.18 $\mu$m.}
\end{figure}

A measurement of the dark current was subtracted from all frames. A
measurement of the inside of the telescope dome was used to correct for
pixel-to-pixel variations in the detector responsivity. Background subtraction
was performed by subtracting frames with the star at different positions on
the slit. Programme star spectra were divided by standard star spectra to
correct for the severe absorption by telluric OH lines. The spectra were
flux-calibrated by assuming L-band magnitudes of 2.797 and 3.664 mag, and
effective temperatures of 7000 and 8000 K for HR0077 and HR2015, respectively.
To give an idea of the importance of telluric absorption in the L-band, we
show the composite spectrum of HR2015 in Fig.\ 1. We overplot a blackbody of
8000 K for comparison. The spectra do not match perfectly around 3.18 $\mu$m,
because the instrument is more sensitive at the short- than at the
long-wavelength end. Near 3.10 $\mu$m telluric absorption often reduces the
measured intensity of programme stars to very low levels, introducing
relatively large uncertainties. Wavelength calibration was performed by a
linear fit to the positions of the many telluric lines.

%
%
\begin{figure*}[tb]
\centerline{\hbox{
\psfig{figure=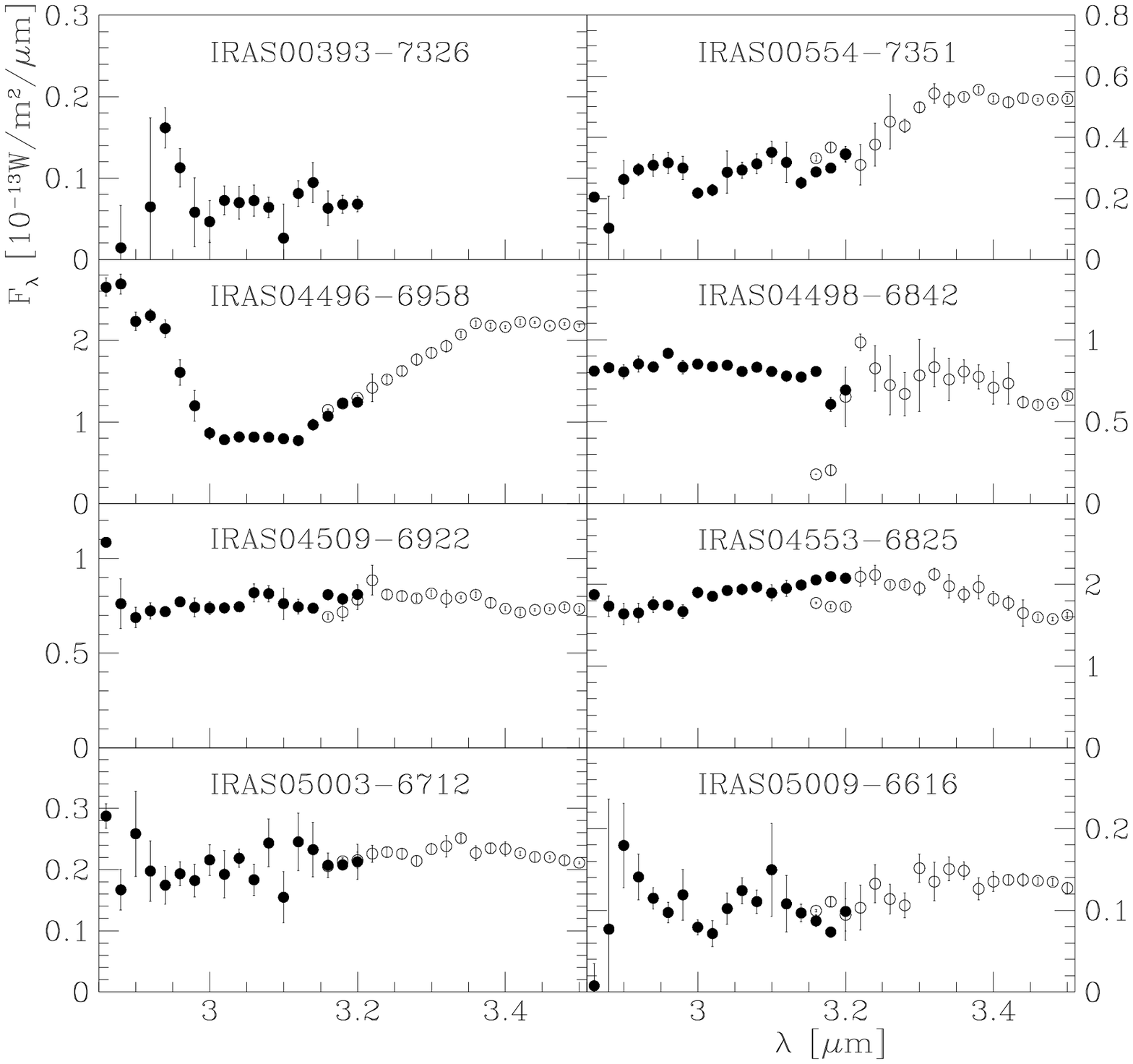,width=90mm}
\psfig{figure=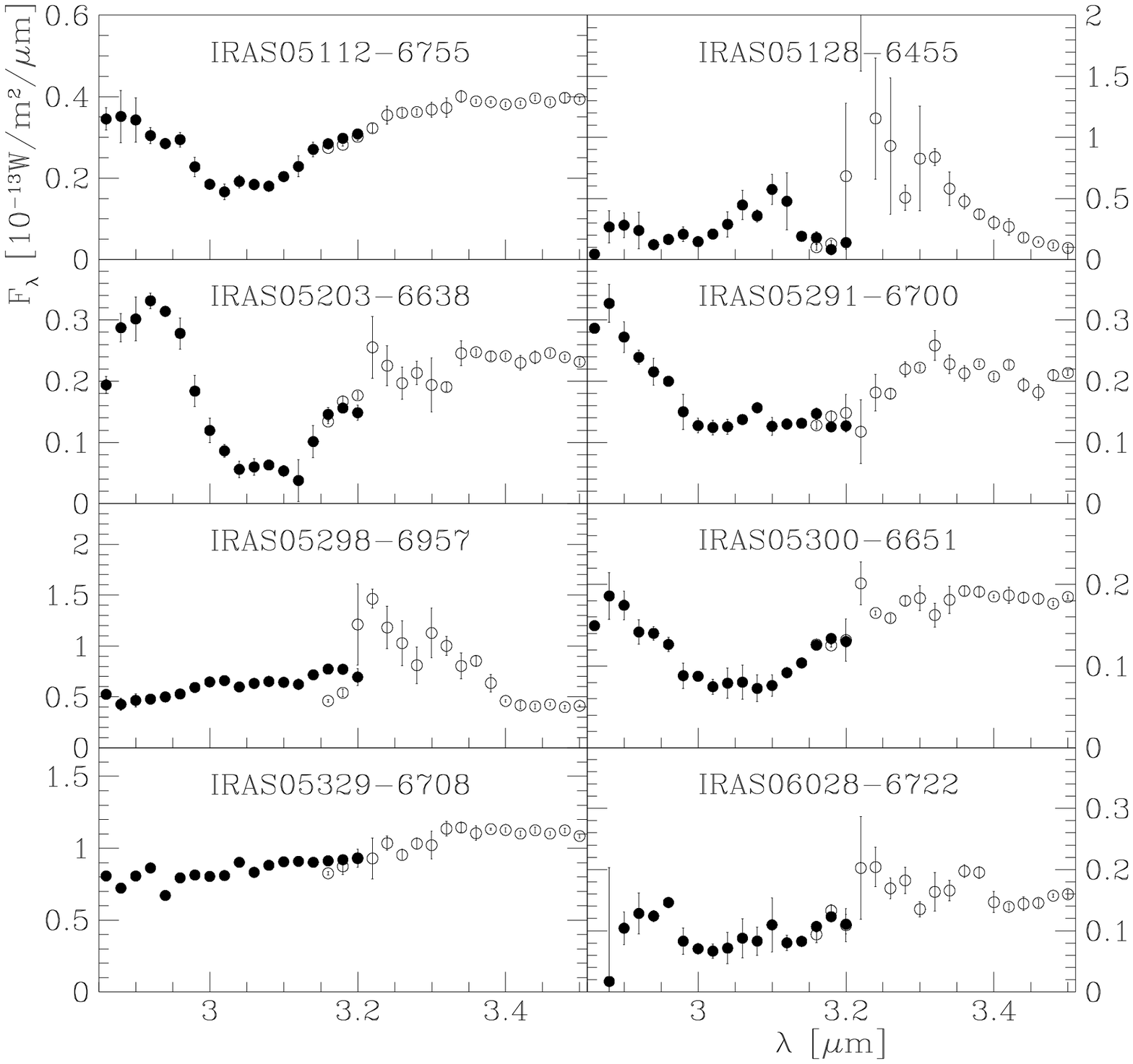,width=90mm}
}}
\caption[]{3 $\mu$m Spectra of obscured Asymptotic Giant Branch stars in the
Magellanic Clouds (plus LMC red supergiant IRAS04553$-$6825). Absorption
between 3.0 and 3.2 $\mu$m is due to HCN and C$_2$H$_2$ molecules in the
extended atmospheres of carbon stars (e.g.\ IRAS04496$-$6958). Oxygen-rich
(M-type) stars display a featureless continuum (e.g.\ IRAS04509$-$6922).}
\end{figure*}

We rejected pixels with flux densities more than 3-$\sigma$ off the median
within a running bin of 22 pixels width. The spectra were rebinned to 0.02
$\mu$m steps reducing the effective spectral resolving power to $R\sim100$. We
assigned errorbars that we derived from the spread of pixel values within the
0.02 $\mu$m bins. The long-wavelength part of the spectrum of IRAS05298$-$6957
is of low quality because we could not guide anymore in the morning twilight.
The 3 $\mu$m spectra are shown in Fig.\ 2.

\section{Results}

\subsection{Classification into carbon and M stars}

Both SMC stars in our sample have IR colours indicative of carbon-rich dust
(Groenewegen \& Blommaert 1998). The drop in flux density in the spectrum of
IRAS00393$-$7326 going from 2.94 $\mu$m to beyond 3.0 $\mu$m suggests indeed
absorption by HCN and C$_2$H$_2$. The overall shape of the spectrum of
IRAS00554$-$7351 suggests moderate absorption on a red continuum. An
unpublished 3 $\mu$m spectrum of this star taken by Whitelock also confirms
its carbon-rich nature (as cited by Wood et al.\ 1992).

IRAS04509$-$6922 and IRAS04553$-$6825 are known to have oxygen-rich
photospheres from their optical spectral types: M10 (paper IV) and M7.5 (Elias
et al.\ 1986), respectively. Maser emission from oxygen-rich molecules has
been observed from IRAS04553$-$6825 (Wood et al.\ 1986; van Loon et al.\ 1996,
1998b), IRAS05298$-$6957 (Wood et al.\ 1992), and IRAS05329$-$6708 (Wood et
al.\ 1992), indicating oxygen-rich CSEs. Their 3 $\mu$m spectra show a
featureless continuum. IRAS04498$-$6842 and IRAS05003$-$6712 also have a
featureless 3 $\mu$m spectrum and hence are M stars, in agreement with their
IR colours (paper IV). The curvature of the spectrum of IRAS04553$-$6825 hints
at absorption by the wings of H$_2$O vapour (band-head at 2.7 $\mu$m).

The other LMC stars are carbon stars. IRAS04496$-$6958 is the showcase of a
carbon star spectrum in our sample. IRAS05203$-$6638 and IRAS05291$-$6700 also
have carbon-rich CSEs according to their IR colours (paper IV), and they
display very strong 3 $\mu$m absorption indeed. IR colours are not conclusive
for IRAS05009$-$6616 (paper IV), IRAS05300$-$6651, and IRAS06028$-$6722 (paper
III), but our 3 $\mu$m spectra suggest they are carbon stars. IRAS05128$-$6455
cannot be classified easily. Its IR colours are inconclusive (paper IV). The
quality of its 3 $\mu$m spectrum is low, but the lack of flux between 3.15 and
3.20 $\mu$m, where telluric extinction is relatively weak (see Fig.\ 1),
suggests absorption when compared to the continuum level around 3.35 $\mu$m.
IRAS05112$-$6755 is undoubtedly a carbon star, as seen from the 3 $\mu$m
spectrum. This overrules our previous and somewhat ambiguous classification as
oxygen-rich (papers II \& IV), explains our non-detection of OH maser emission
(paper IV), and is in much better agreement with the highly evolved state of
this AGB star (Wood 1998).

\subsection{Strength of 3 $\mu$m absorption}

%
%
\begin{table}
\caption[]{Equivalent widths of the 3 $\mu$m absorption, with 2-$\sigma$ error
estimates. Also listed are the bolometric magnitudes, $(K-L)$ colours, and
chemical types.}
\begin{tabular}{lllll}
\hline\hline
IRAS         & $M_{\rm bol}$ & $(K-L)$ & $W_{3{\mu}m}$            & chemistry \\
\hline
00393$-$7326 & $-5.9$        & 2.49    & ?                        & (carbon) \\
00554$-$7351 & $-6.4$        & 2.55    &           $0.070\pm0.03$ & (carbon) \\
04496$-$6958 & $-7.0$        & 1.53    &           $0.176\pm0.02$ &  carbon  \\
04498$-$6842 & $-8.4$        & 0.89    &           $0.011\pm0.04$ &  oxygen  \\
04509$-$6922 & $-7.4$        & 1.24    & \llap{$-$}$0.008\pm0.02$ &  oxygen  \\
04553$-$6825 & $-9.5$        & 1.91    & \llap{$-$}$0.023\pm0.02$ &  oxygen  \\
05003$-$6712 & $-5.6$        & 1.60    &           $0.009\pm0.04$ &  oxygen  \\
05009$-$6616 & $-6.1$        & 2.40    &           $0.107\pm0.06$ & (carbon) \\
05112$-$6755 & $-6.2$        & 3.40    &           $0.102\pm0.03$ &  carbon  \\
05128$-$6455 & $-6.5$        & 1.96    &           $0.091\pm0.08$ & (carbon) \\
05203$-$6638 & $-5.7$        & 1.80    &           $0.166\pm0.02$ &  carbon  \\
05291$-$6700 & $-5.6$        & 1.30    &           $0.143\pm0.02$ &  carbon  \\
05298$-$6957 & $-7.8$        & 2.76    &           $0.004\pm0.05$ &  oxygen  \\
05300$-$6651 & $-5.1$        & 2.57    &           $0.129\pm0.03$ &  carbon  \\
05329$-$6708 & $-7.0$        & 2.10    &           $0.007\pm0.04$ &  oxygen  \\
06028$-$6722 & $-6.4$        & 3.10    &           $0.089\pm0.06$ & (carbon) \\
\hline
\end{tabular}
\end{table}

We have determined the equivalent width $W$ of the 3 $\mu$m feature. The
continuum can be estimated near 2.9 and 3.3 $\mu$m, which is also the major
source of uncertainty. Errors (2-$\sigma$) were estimated by measuring $W$ for
the highest and lowest continuum levels that are still compatible with the
data. The results are listed in Table 2, together with the $(K-L)$ colours.
The SMC photometry is from Groenewegen \& Blommaert (1998).

%
%
\begin{figure}[tb]
\centerline{\psfig{figure=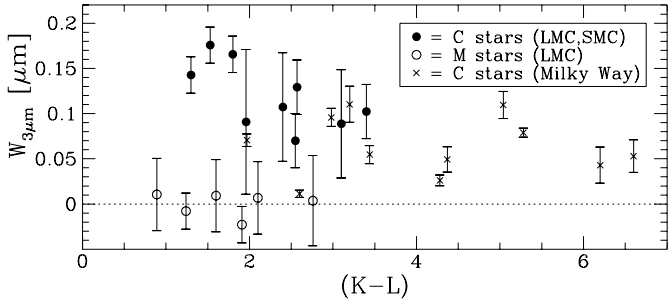,width=88mm}}
\caption[]{Equivalent width of the 3 $\mu$m molecular absorption feature
versus the $(K-L)$ colour for the carbon stars and M-type stars in our sample,
and the galactic carbon star sample from Groenewegen et al.\ (1994). Errorbars
are $\pm2$-$\sigma$. Carbon stars have absorption, M-type stars do not. The
absorption in carbon stars decreases with redder $(K-L)$ due to the increase
in circumstellar emission.}
\end{figure}

We plot the $W_{3{\mu}m}$ versus the $(K-L)$ colours (Fig.\ 3). The scatter of
the M-type stars around $W_{3{\mu}m}=0$ $\mu$m gives a good impression of the
accuracy of the estimates of $W_{3{\mu}m}$. All carbon stars are well
separated from the oxygen-rich M-type stars. Despite the larger than typical
error on $W_{3{\mu}m}$ for IRAS05128$-$6455, it is evident that it is a carbon
star. We also plot equivalent widths that we measured for the galactic carbon
stars from Groenewegen et al.\ (1994).

In the LMC, the $(K-L)$ colours of the carbon stars are redder than those of
the M-type stars. As we could only take spectra of stars that were
sufficiently bright in the L-band, bolometrically faint stars only entered our
sample if they had red $(K-L)$ colours. The carbon stars in our sample are, on
average, somewhat less luminous than the M stars. As the inner radius of the
CSE scales with luminosity, at a given mass-loss rate, the CSEs of the carbon
stars are optically thicker than the CSEs of the M stars.

The three bluest carbon stars show the strongest absorption feature. The
absorption is formed predominantly within the dust-free inner region of the
CSE. The redder carbon stars have stronger dust continuum emission, decreasing
the contrast between the absorption feature and the continuum (Ridgway et al.\
1983). The dust continuum emission becomes important for sources with dust
optical depths corresponding to $(K-L)\gsim2$ mag. This trend is prominent in
our data of Magellanic Cloud stars, and matches well with the Milky Way stars
of Groenewegen et al.\ 1994 (see also Noguchi et al.\ 1981, 1991).

The 3 $\mu$m absorption in the SMC source IRAS00554$-$7351 is relatively weak.
Note that the galactic carbon star IRAS21377$+$5042 has $(K-L)\sim2.6$ mag and
$W_{3{\mu}m}\sim0.01$ $\mu$m, which would make it indistinguishable from
M-type stars with the typical signal-to-noise in our data. However, it has
very blue IR colours and no molecular emission could be detected at mm
wavelengths (Groenewegen et al.\ 1994), which makes it a highly peculiar
object. It is unlikely to encounter such a star in our sample of stars.

\subsection{Luminous carbon stars}

%
%
\begin{figure}[tb]
\centerline{\psfig{figure=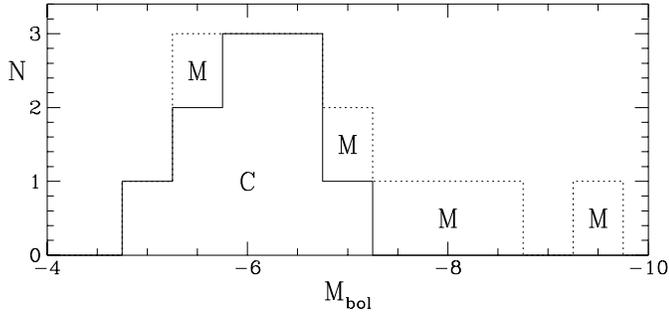,width=88mm}}
\caption[]{Histogram of the luminosity distribution of carbon stars and M
stars. Carbon stars populate the AGB up to the highest luminosities. Luminous
M stars are probably red supergiants.}
\end{figure}

In Table 2 we also list the bolometric magnitudes. We determined bolometric
magnitudes for the SMC stars in the same way as for the LMC stars (papers II,
III \& IV) by spline fitting to the spectro-photometric energy distribution
(see Whitelock et al.\ 1994). The accuracy of this method is $\sim0.1$ mag.
The near-IR photometry are single-epoch data, and time-averaged luminosities
may differ from the values presented here. Bolometric amplitudes of galactic
carbon stars are $\sim0.6$ mag (Le Bertre 1992) and probably similar to those
of galactic OH/IR stars (Le Bertre 1993), hence we expect single-epoch
luminosities to be within 0.3 mag of the time-averaged luminosity. We adopt
distance moduli of 18.7 and 19.1 mag for the LMC and SMC (Feast 1998). These
constitute a rather long distance scale, with the average estimate for the
distance moduli for the LMC and SMC being 18.55 and 18.97 mag, respectively
(Walker 1998).

In Fig.\ 4 we plot a histogram of the luminosities of the carbon stars and M
stars in our sample. We note that due to the small number of stars, and a
possible bias towards carbon stars when we selected the stars, one should not
over-interpret these histograms as the luminosity functions of obscured M-type
and carbon stars in the MCs. However, there is no indication for a deficit in
carbon stars brighter than $M_{\rm bol}=-6$ mag: carbon stars do exist up to
the highest AGB luminosities: $M_{\rm bol}\sim-7$ mag. IRAS04498$-$6842 and
IRAS05298$-$6957 may be luminosity enhanced due to the effects of HBB, or they
may actually be red supergiants.

\section{Discussion}

\subsection{Luminous carbon stars}

Luminous carbon stars may be formed after the AGB star has become obscured by
its dusty CSE. When mass loss has reduced the stellar mantle below a critical
mass, HBB switches off (Boothroyd \& Sackmann 1992). If such a star
experiences another thermal pulse before leaving the AGB, it may become a
carbon star, after all (Frost et al.\ 1998; Marigo et al.\ 1998). This
scenario is consistent with our spectroscopic confirmation of luminous carbon
star candidates that are dust-enshrouded as a result of their AGB mass loss.
The most luminous of these, IRAS04496$-$6958, has an estimated $M_{\rm
bol}=-7$ mag. However, we note that this object is peculiar in that it has a
(minor) oxygen-rich dust component in its CSE (Trams et al.\ 1999a,b).

As the most luminous carbon star in the LMC is not experiencing HBB, its
luminosity must not exceed the classical limit to the AGB luminosity: $M_{\rm
bol}\gsim-7.1$ mag (Wood et al.\ 1983; Boothroyd \& Sackmann 1992). This sets
an upper limit to the distance modulus of the LMC of 18.8 mag.

\subsection{Carbon and nitrogen enrichment}

Differences in metallicity may be expected to affect the molecular and dust
abundances in the CSEs. Yet carbon stars in the LMC and the Milky Way appear
to follow the same sequence of $W_{3{\mu}m}$ versus $(K-L)$. We here
investigate whether this can be understood.

The observed $W_{\rm observed}$ is to a good approximation a dilution factor
$(1+\xi)$ smaller than the $W_\star$ in the purely photospheric spectrum:
\begin{equation}
W_{\rm observed} = \int \left( \frac{ f_{\lambda,{\rm continuum}} - f_\lambda
}{ f_{\lambda,{\rm continuum}} } \right) {\rm d}\lambda = \frac{ W_{\star} }{
1+\xi }
\end{equation}
where the diluted flux $f$ is given by:
\begin{equation}
f = F {\rm e}^{-\tau} + S \tau {\rm e}^{-\tau}
\end{equation}
with the photospheric flux $F$, dust emission source function $S$, and optical
depth $\tau$. We see that $\xi$ is proportional to $\tau$ (times the ratio of
$S/F_{\rm continuum}$). The difference between the observed $(K-L)$ and the
photospheric $(K-L)_0$ colour is also proportional to $\tau$. Hence, for a
given $W_\star$, when the metallicity decreases, so does the optical depth of
the CSE. The $(K-L)$ colour becomes bluer, and the carbon star shifts along
--- not away from --- the $W_{3{\mu}m}$ versus $(K-L)$ sequence. The fact that
carbon stars in the LMC and the Milky Way appear to follow the same sequence
of $W_{\rm observed}$ versus $(K-L)$ therefore implies that $W_\star$ is
independent of metallicity. As the 3 $\mu$m absorption is caused by HCN and
C$_2$H$_2$ molecules, we now investigate how their abundances are expected to
depend on metallicity.

As C$_2$H$_2$ associates at lower temperatures and gas pressures than HCN
(Ridgway et al.\ 1978), free carbon is expected to be preferentially locked
into HCN. We assume that the number of carbon atoms that are left after all
oxygen is consumed in the formation of CO molecules, the free carbon abundance
$N$(C$-$O), will combine with the available nitrogen atoms $N$(N) to form HCN
molecules. The free carbon abundance is determined by the initial free carbon
abundance, $N_0$(C$-$O), and the production and dredge-up of carbon during the
thermal-pulsing AGB, $\delta$(C). Likewise, the nitrogen abundance is
determined by the initial nitrogen abundance, $N_0$(N), and the production and
dredge-up of nitrogen, $\delta$(N). The latter takes place already before the
thermal-pulsing AGB, but is especially important when HBB occurs. The initial
abundance is proportional to the metallicity $Z$. Both the free carbon and the
nitrogen abundance can be written in identical form:
\begin{equation}
N = \alpha Z + \delta
\end{equation}
It is not well known what is the metallicity dependence of $\delta$. With
$Z_{\rm LMC} = \frac{1}{2} \times Z_{\rm MW}$, the HCN abundance ratio between
a carbon star in the LMC and a corresponding carbon star in the Milky Way (MW)
can be written as
\begin{equation}
\frac{N_{\rm LMC}({\rm HCN})}{N_{\rm MW}({\rm HCN})} =
\frac{0.5+\eta_{\rm LMC}}{1+\eta_{\rm MW}}
\end{equation}
where the enrichment term $\eta$ is given by
\begin{equation}
\eta = \frac{\delta}{\alpha Z_{\rm MW}}
\end{equation}

Initially there is more oxygen than carbon, and $\alpha$(C$-$O) is negative.
To make a carbon star, $\delta$(C) must exceed
\mbox{$\mid\alpha$(C$-$O)$Z\mid$}. With the same value for $\eta$(C)$<-1$ in
the LMC and the Milky Way, Eq.\ 4 implies that more HCN would be produced in
the LMC than in the Milky Way if the free carbon abundance limits the HCN
abundance, much in the same way that it is easier to make carbon stars at
lower metallicity because there is less oxygen to bind into CO. If, however,
the nitrogen abundance limits the HCN abundance, then, as $\alpha$(N) and thus
$\eta$(N) are positive, Eq.\ 4 implies that less HCN will be produced in the
LMC than in the Milky Way unless the nitrogen enrichment term $\eta$ is larger
at lower metallicity. The question now is which species is more abundant in
magellanic carbon stars: nitrogen or free carbon?

The photospheric abundance of a carbon star near the end of its AGB life
resembles that of the PN that will be ejected. Of the $16+15$ PNe in the
LMC$+$SMC for which Leisy \& Dennefeld (1996) measured abundances, $8+10$ have
$N$(C$-$O)$>0$, i.e.\ their progenitors left the AGB as carbon stars. All of
these have $N$(C$-$O)$>N$(N), the only exception being SMP47 in the LMC of
Type I (nitrogen-enhanced). This would suggest that in the MCs, $N$(HCN) in
carbon stars is limited by $N$(N), not by $N$(C$-$O). However, the PNe in the
MCs are all less luminous than about 10,000 L$_\odot$, corresponding to the
luminosity of the faintest carbon star in our sample. The only more luminous
PN in the LMC is SMP83, which is oxygen-rich (Dopita et al.\ 1993). In
luminous AGB stars with massive cores, HBB enhances the nitrogen abundance.
This starts to be effective around $M_{\rm bol}=-6$ mag, where the optical
luminosity function of carbon stars in the MCs truncates. However, when mass
loss continues, HBB diminishes. Stellar evolution codes show the N/O to
decrease again, whilst the C/O increases (Marigo et al.\ 1998). If an AGB star
becomes a carbon star only after having experienced HBB, nitrogen may not be
as over-abundant as observed for instance in PNe of Type I --- that are
usually oxygen-rich. So we cannot be sure whether our sample of magellanic
carbon stars also have $N$(C$-$O)$>N$(N).

Anyway, any surplus of carbon ends up in C$_2$H$_2$. Thus, all carbon
contributes to the 3 $\mu$m absorption, either through HCN or C$_2$H$_2$. If
this is correct, then the strength of the 3 $\mu$m absorption is determined by
the free carbon abundance, no matter what the nitrogen abundance is. Hence, in
analogy to Eq.\ 4 we expect the 3 $\mu$m absorption to be stronger in the LMC
than in the Milky Way, in disagreement with our observations. If the carbon
enrichment is much larger than the initial carbon abundance ($\eta$(C)$\ll-1$)
the differences in 3 $\mu$m absorption strength may be too small to appreciate
in our data. Alternatively, the carbon enrichment may be less at lower
metallicity. Another explanation may lie in the fact that stellar photospheres
are warmer at lower metallicity. The rotational bands of HCN and C$_2$H$_2$
are excited at temperatures between $\sim1000$ and 1500 K (Ridgway et al.\
1983), which occurs more distant to a warmer stellar photosphere. The 3 $\mu$m
band strength may be weakened by this effect, canceling the expected increase
in strength due to the more abundant free carbon at lower metallicity.

\acknowledgements{We thank Dr.\ Brooke Gregory for assistance during the
observations, Dr.\ Peter Wood for sharing results prior to publication, Dr.\
Rens Waters for critical reading of an earlier version of this manuscript, and
the referee, Dr.\ Shaun Hughes, for suggestions that helped improve the
paper.\\ Mais que muitos obrigadinhos para o anjinho Joana --- Yes, I know.}

\end{document}